# 九论以用户为中心的设计：智能时代的“用户体验 3.0”范式

许 为*

（浙江大学 心理科学研究中心，杭州 310058）

**研究要点**

1. 梳理了用户体验（UX）范式的跨时代演进和智能时代对 UX 范式的新要求。
2. 首次提出了智能时代的“UX 3.0”范式框架和方法体系。
3. 展望了“UX 3.0”范式今后的研究和应用。

**摘要**

基于“以用户为中心设计”理念的用户体验（UX）领域正在迈向智能时代，但是现有的 UX 范式主要针对非智能系统，缺乏针对智能系统 UX 的系统化方法。纵观 UX 的发展历程，UX 范式呈现出跨技术时代的演进特征。当前，智能时代对 UX 范式提出了新要求。为此，本文从 UX 范式出发，提出智能时代的“UX 3.0”范式框架以及相应的方法体系。“UX 3.0”范式框架包括五大类 UX 方法：生态化体验、创新赋能体验、AI 赋能体验、人智交互体验、人智协同合作体验方法，其中每一类方法都包括相应的 UX 范式取向。“UX 3.0”范式的提出有助于提升现有 UX 方法，为智能系统 UX 研究和应用提供方法论支持。最后，本文展望 “UX 3.0” 范式今后的研究和应用。

***作者**：许为，男，博士，研究员，e-mail: xuwei11@zju.edu.cn

# User-Centered Design（IX）：
# A "User Experience 3.0" Paradigm Framework in the Intelligence Era

XU Wei

(Center for Psychological Sciences, Zhejiang University,
Hangzhou 310058, China)

**Abstract**

The field of user experience (UX) based on the design philosophy of "user-centered design" is moving towards the intelligence era. Still, the existing UX paradigm mainly aims at non-intelligent systems and lacks a systematic approach to UX for intelligent systems. Throughout the development of UX, the UX paradigm shows the evolution characteristics across technological eras. The intelligence era has put forward new demands on the UX paradigm. For this reason, this paper proposes a "UX 3.0" paradigm framework and the corresponding UX methodology system in the intelligence era. The "UX 3.0" paradigm framework includes five categories of UX methods: ecological experience, innovation-enabled experience, AI-enabled experience, human-AI interaction-based experience, and human-AI collaboration-based experience methods, each providing for corresponding UX paradigmatic orientations. The "UX 3.0" paradigm helps enhance existing UX methods and provides methodological support for the research and applications of UX in developing intelligent systems. Finally, this paper looks forward to future research and applications of the "UX 3.0" paradigm.

**Key words:** User-centered design, user experience, user experience paradigm and method, human-AI interaction, human-AI collaboration

# 1 引言

Norman (1986)提出的用户体验（user experience，UX）概念和“以用户为中心设计”理念带动了计算机时代 UX 和人机交互(HCI)新兴领域的发展(Nielson, 1993; 许为，2003，2005)。近二十年，中国 UX 领域进入繁荣阶段，从可用性设计和人机交互，到整体 UX、体验设计、服务体验、体验思维和体验管理等（董建明，傅利民，饶培伦等，2021；辛向阳，2019；黄峰,赖祖杰, 2020；黄峰,黄胜山,苏志国,2022；许为，2017, 2019a）。目前，UX 开始迈进智能时代，智能技术和智能产品的普及给 UX 的进一步发展带来了新要求以及新机遇(许为，2019b，2020a；许为，高在峰, 葛列众, 2022)。

本文从 UX 研究和应用方法论的角度出发，主要回答以下问题：智能时代 UX 应该采用什么范式来进一步推动 UX 研究和应用？为回答此问题，在梳理 UX 范式跨时代演进和分析智能时代 UX 范式面临的新要求基础上, 根据我们以往的工作，本文提出一个智能时代的“UX 3.0”范式框架，并详细论述该框架包括的 UX 方法体系，最后展望“UX 3.0”范式今后的工作。目的是为智能时代开展更加有效的 UX 研究和应用提供人因科学方法论的支持。

## 2 智能时代对 UX 范式的新要求

### 2.1 UX 的跨时代演进

回顾 UX 发展历史，根据技术平台、应用领域、用户需求、人机交互等特征，UX 的发展可以初步被划分为三个阶段（见表 1）(许为, 2019a，修改版)。“个人电脑/互联网时代”开创了 UX 的第一阶段。2007 年，苹果公司推出的 iPhone 创新了智能手机人机交互方式和移动体验，促进了基于移动互联网技术应用和创新发展，标志着 “移动互联网时代”的开始, 同时推动了 UX 第二阶段的发展。2015 年，谷歌公司 AlphaGo 人工智能( AI) 产品问世，深度机器学习方法等技术全面展开, 标志着基于 AI、机器学习( ML) 、大数据、云计算等技术的智能时代的来临，各种智能系统产品开始进入人们的日常工作和生活，这标志着 UX 第三阶段的开始。

表 1　UX 发展的跨时代演进特征

| 演进特征 | 第一阶段：UX 1.0<br>（1980 年代后期 - ～2007 年） | 第二阶段：UX 2.0<br>（～2007 年 - ～2015 年） | 第三阶段：UX 3.0<br>（～2015 年 - ） |
|---|---|---|---|
| 技术时代 | 个人电脑/互联网时代 | 移动互联网时代 | 智能时代 |
| 技术平台 | 个人电脑,互联网等 | + 移动互联网,智能手机，平板电脑等 | + AI, 深度机器学习，大数据,云计算,5G 网络,区块链等 |
| 应用领域 | 互联网网站，电商零售，个人电脑应用等 | + 移动互联网，消费/商业互联网, APP 等 | + 垂直智能行业(医疗、家居、交通、制造等）, 智能物联网/工业互联网,机器人,智能驾驶车,虚拟现实 (VR/AR/MR) + 元宇宙等 |
| **UX 理念** | 以用户为中心 | 以用户为中心 | 以用户为中心（包括“以人为中心 AI”）（许为，2019b） |
| **用户需求** | 产品功能性,可用性等 | + 整体 UX(开始)（董建明，傅利民，饶培伦等,2016），信息安全等 | 整体 UX + 人机交互智能化、自然化、个性化、情感化， 人智协同合作，伦理道德， 个人隐私，参与感，决策自主权，技能成长等 |
| **UX 范围** | 单一产品人机界面的可用性 | UX。UX 不局限于人机界面可用性，也包括业务流程等用户交互接触点（许为，2017） | (开始)生态化、系统化、创新式、智能化、协同合作式 UX。详见以下讨论 |
| **UX 跨度** | 产品开发阶段 | 产品全生命周期（开发前、开发、开发后） | 产品全生命周期 + 智能生态系统及智能社会技术系统的宏观环境（许为，2022c） |
| **人机交互** | 图形化， 显式化，用户精准输入等 | + 触摸式，多模态等 | + 自然化(语音，体感交互等)，智能化（用户模糊输入、用户意图推理、机器主动式等）,普适化，隐式化，虚拟化等（许为，2022a） |

**注：**表中“+”是指增加的特征。

如表 1 所列，根据技术时代所划分的 UX 三个阶段各自表现出明显的阶段特征。从 UX 角度看，根据用户需求，第一阶段的 UX 研究和应用主要局限在产品开发期间，围绕单一交互产品中相对简单的人机界面可用性设计和测试等活动展开。进入第二阶段，UX 研究和应用开始向整体 UX 拓展，除了针对人机界面的可用性设计，UX 专业人员也开始对业务流程等各类用户交互接触点进行体验优化，设计跨度开始覆盖产品全生命周期。进入第三阶段，随着更加丰富化用户需求的涌现，人机交互特征更加丰富和智能化，系统化的 UX 设计思维开始形成，在更大范围和跨度内，围绕智能人机交互以及人智协同合作的体验设计成为 UX 的重点，这些新特征也对 UX 研究和应用提出了新挑战。

### 2.2 智能时代对 UX 方法的新要求

基于 AI 技术的智能产品正在造福人类，但是，研究表明不恰当的技术开发可能影响 UX，甚至伤害人类。AI 事故数据库已经收集了 1000 多起事故,这些事故包括自动驾驶汽车撞死行人、交易算法错误导致市场“闪崩”等，这些事故生也必然导致用户的负面体验 (Mcgregor, 2023)。许多 AI 技术人员在开发中基本按照“以技术为中心”的理念，认为以往人机交互无法解决的问题目前已被智能技术解决（如语音输入）,因此人机交互和界面设计不必再过多地考虑 UX；而 UX 专业人员往往在 AI 项目产品需求定义后才参与项目，限制了他们对智能系统设计的影响，导致一些 AI 项目失败（Yang et al., 2020; Budiu & Laubheimer, 2018)。

为应对这些状况，研究者提出了“以人为中心 AI”的理念（Li, 2018; 许为，2019b；Xu，2019; Shneiderman, 2020）。“以人为中心 AI”理念的一个重要内容就是强调优化智能系统的人机（智）

交互和体验设计，解决存在的一系列人因（human factors）问题。UX 对智能系统的成败起着重要作用，例如，智能人机界面的体验设计，迭代式的智能界面原型化和测试等。

一个领域的范式决定了该领域研究和应用的取向、重点及方法(许为，高在峰, 葛列众, 2022)。纵观 UX 发展阶段，UX 范式呈现出明显的阶段性演进特征；同时，借助于新兴技术, UX 范式一直在发展，这种发展提升了 UX 研究和应用的方法，也推动了 UX 领域的不断发展。

如表 1 所示，在 UX 发展的第一阶段，UX 范式主要基于针对人机界面设计和测试的可用性工程方法（Neilson,1993）。在第二阶段，UX 范式开始超越人机界面的可用性，一些拓展的 UX 取向和方法开始应用。例如，整体（全部）UX（董建明，傅利民，饶培伦等，2016），增强型 UX 方法（Finstad, K., Xu et al., 2009; Xu, 2014; 许为，2017）。进入智能时代的 UX 第三阶段，UX 领域必然需要合适的范式，表 2 概括了智能时代对 UX 范式和方法的新要求。

表 2　智能时代对 UX 范式和方法的新要求

| 分类 | 对 UX 研究和应用的新要求 | 对 UX 方法的新要求 | UX 新目标 |
|---|---|---|---|
| **用户新需求** | • 智能化、自然化、个性化、情感化的人机（智）交互<br>• 维护人类伦理道德，个人数据隐私权，社会公平感等<br>• 人-智能系统之间的协同合作，人类智能与机器智能互补，人类智能增强<br>• 用户参与感，决策自主权，个人技能成长<br>• 提升工作生活质量 | • 支持不同用户群体、同一用户群体内不同层次的体验需求<br>• 利用人类智能与机器智能的差异性和互补性，提升人机系统的整体绩效，获取最佳体验， | 满足用户多层次需求，提供更丰富的 UX |
| **生态化体验**（参见 3.2.1） | • 跨产品全生命周期（即市场调研、产品创新、品牌设计、产品开发、市场投放、用户支持等各个环节）的整体体验 (Xu, 2012)<br>• 跨技术平台和设备、跨服务、跨内容的无缝交互体验<br>• 跨系统前端（如 UI）、中端（如业务逻辑和流程）、后端（如数据库）的整体体验<br>• 在智能社会技术系统的宏观生态环境中，优化智能社会形态（如智能交通、智能医疗）中技术与社会之系统（如组织、文化）间的交互（许为，2022c） | • 包含宏观、横向、系统化的设计思维<br>• 系统化的 UX 方法<br>• 采纳跨学科方法<br>• 全体验生态的整体 UX 设计 | 优化全体验生态系统，提供无缝衔接、端到端的整体体验 |
| **创新赋能体验**（参见 3.2.2） | • 市场竞争对体验设计的新要求（许为，2019a）<br>• 快速产品迭代速度导致设计空间受限和设计趋同性<br>• 用户期望提升，例如，差异化体验、实时数据（用户行为、上下文场景等）体验、基于用户痛点的创新体验<br>• 基于智能新技术（如 AI,大数据，智能人机交互）来提升用户体验 | • 利用新技术（如 AI、大数据、智能人机交互），提升传统 UX 方法（谭浩等，2020）<br>• 基于创新理念的系统化、体验驱动式创新设计<br>• 具操作性的 UX 创新方法 | 通过体验驱动式创新，UX（以及专业人员）能够主动地主导产品设计 |
| **AI 赋能体验**（参见 3.2.3） | • 利用智能新技术，提升 UX 活动效率（用户研究、界面原型化、体验测试等）（Xu，2023）<br>• 智能感知用户实时在线行为，智能推理用户意图和需求<br>• 智能洞察用户个性化需求，推送个性化功能和内容<br>• 智能分析和推理用户大数据（尤其是质化数据）<br>• 支持人智协同人机界面设计<br>• 实时体验支持（如智能推荐系统、智能语音支持助手） | • 高效的数智化 UX 数据生成方法（如数智化用户画像、数智化客户旅程图）(Salminen et al., 2021)<br>• 基于“AI 作为设计新材料”理念的体验设计 (Dove et al., 2017)<br>• 基于 AI 生成法的 UI 原型化(Li et al., 2020)<br>• 基于 UX 与 AI 专业人员合作的跨学科方法 | 提升 UX 方法和设计效率，优化体验设计 |
| **人智交互体验**（参见 3.2.4）） | • 基于人智交互的数智化体验(许为，2022a)<br>• 可视化、可解释的智能系统人机界面（杨强等， 2022）<br>• 自然式、多通道融合的智能人机交互<br>• 对用户状态（情景意识、生理、情感、用户意图等）的智能识别和推演<br>• 不确定性条件下（用户模糊输入、“情境化”输入等）对用户交互意图的推理<br>• 社会和情感交互 | • 基于智能人机交互的界面设计新范式和新隐喻<br>• 用户状态和意图的交互和认知建模方法<br>• 支持智能人机交互设计的自然性和有效性<br>• 支持基于社会、情感等交互的体验设计 | 提供更加自然、有效的人机交互 |
| **人智协同合作体验**（参见 3.2.5） | • 基于人机共信、人机分享式态势感知/心理模型/决策和控制的体验 (NASEM, 2021; 许为,葛列众，2020)<br>• 基于人-智组队合作的体验<br>• 基于“人在环路”的人智协同合作式体验<br>• 基于人类可控 AI 的体验 | • 借助工程心理学、人因工程等人因科学方法，提供基于人智组队合作范式的人智协同合作式体验设计(许为，高在峰，葛列众，2022)<br>• 对人智协同合作中团队绩效的体验评价<br>• 基于人机混合智能的人智协同合作体验设计<br>• 基于人智控制分享、人类可控 AI 的体验设计方法（如有意义的人类控制）（许为，葛列众，高在峰，2021) | 利用人机智能互补和协同合作，提升整体人机系统绩效和体验 |

由表 2 所知，从用户新需求、生态化体验、创新赋能体验、AI 赋能体验、人智交互体验以及人智协同合作体验等六个方面，智能时代对 UX 范式和方法提出了一系列新要求，这些新要求比 UX 第二阶段上了一个新的大台阶。同时，智能时代也提升了 UX 新目标（见表 2），这些新目标包括：

- 满足用户多层次需求，提供更丰富的 UX
- 优化全体验生态系统，提供无缝衔接、端到端的整体 UX
- 通过创新体验来主动地主导产品设计
- 提供更加自然、有效的人机交互
- 提升 UX 方法和设计效率，优化体验设计
- 利用人机智能互补和协同合作，进一步提升整体人机系统绩效和体验

## 3 智能时代“UX 3.0”范式

### 3.1 “UX 3.0”范式的概念框架

针对智能时代的 UX，研究者已经开始提出范式和方法论方面的一些初步考虑。例如，研究者利用 AI 和大数据等技术实现对实时在线用户行为等数据的建模，从而为用户提供个性化的设计(谭浩等，2020；Herath & Jayarathne, 2018; Quadrana et al, 2017; 吕超, 朱郑州，2018)，辛向阳(2019)提出了从用户体验转向体验设计的新范式。在人因科学领域，研究者开始关注与智能系统相关的 UX 问题（例如，Stephanidis, Salvendy et al., 2019；董建明，傅利民，饶培伦等，2021；Xu，Dainoff et al., 2021；Shneiderman, et al., 2016; Preece, et al., 2019; 许为, 葛列众, 2020）。但是，目前还没人提出针对智能时代的系统化 UX 范式。

现有的 UX 范式主要基于针对非智能系统的 UX 实践（即第二阶段 UX），缺乏针对智能系统 UX 的系统化方法。这一方面给进入智能时代 UX 实践深水区的 UX 专业人员提出了新挑战; 另一方面，给人因科学提供了提升 UX 范式的新机遇，人因科学（人因工程、工程心理学等）责无旁贷，应该提供学科支持。

为了满足智能时代对 UX 范式的新要求，综合我们以往研究以及他人研究，本文提出一个智能时代 “UX 3.0”范式的概念框架（见图 1）。“UX 3.0”范式定义了由五大类方法组成的 UX 方法体系。

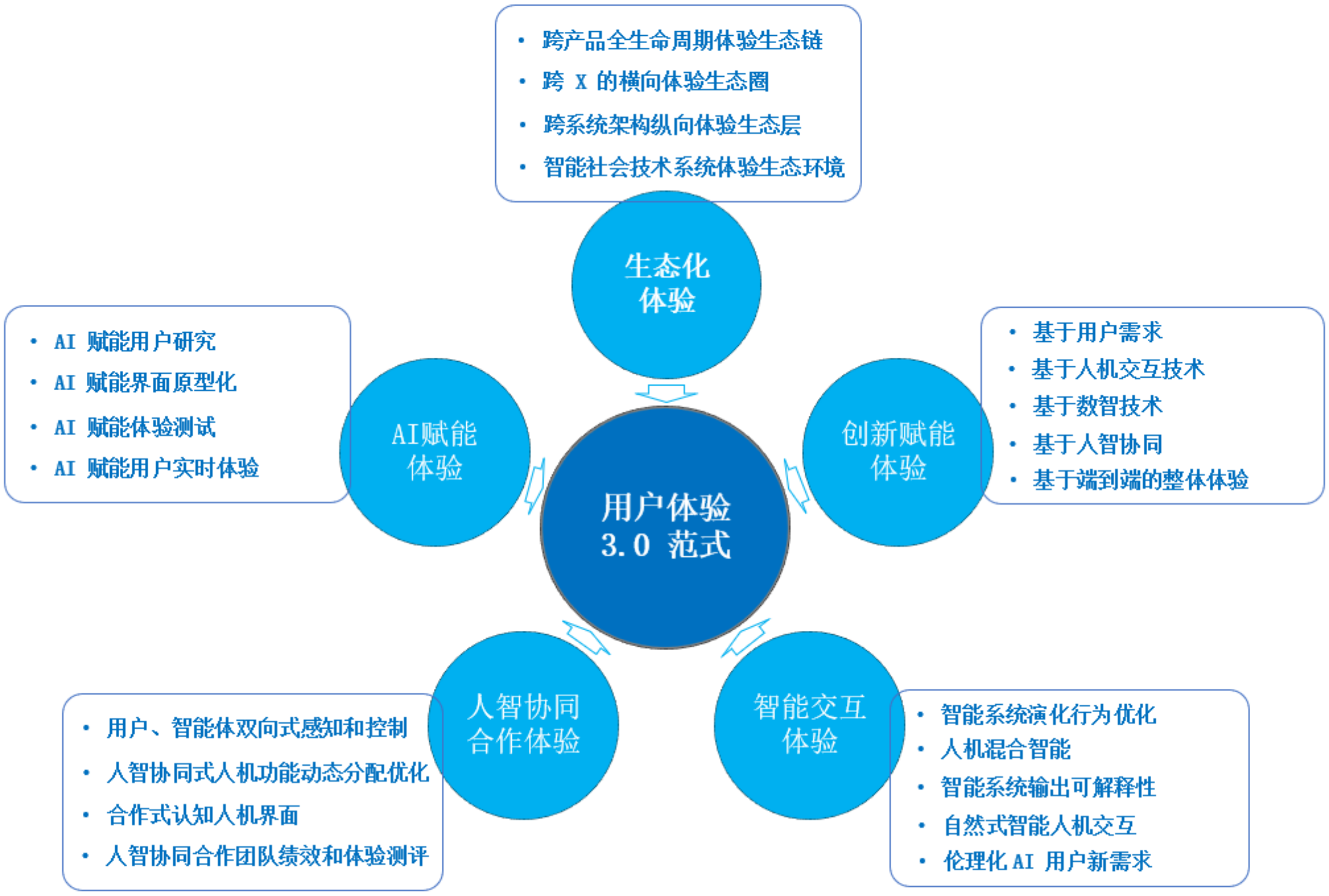

图 1　智能时代“UX 3.0”范式的概念框架

如图 1 所示，“UX 3.0” 范式的概念框架包括五大类 UX 方法：生态化体验、创新赋能体验、AI 赋能体验、人智交互体验、人智协同合作体验。其中，每一类方法都包括了相应的 UX 范式取向（见图 1 中长方形框），例如，创新赋能体验方法是基于用户需求、人机交互技术、数智技术、人智协同以及端到端的整体体验等五个范式取向。这些范式取向 UX 方法提出了具体要求。3.2 节将详细阐述这些范式取向和具体的 UX 方法。

“UX 3.0”范式框架体现了以下基本特征：

- **系统性：**全方位系统考虑 UX。例如，体验的生态化视野，体验驱动创新的视角，系统化的理念；
- **前沿性：**充分考虑智能时代的技术、理念、体验和方法。例如，创新理念，AI 作为 UX 工具（即 AI 赋能体验）、AI 作为人机交互技术（即人智交互体验）、AI 作为人智团队合作队员（即人智协同合作体验）在 UX 中的多重角色，智能时代的人机交互（人智交互）体验，智能时代新型人机关系（人智协同合作）产生的体验；
- **指导性：**定义一系列具体可操作的 UX 方法，从而指导 UX 研究和应用；
- **扩展性：**定义一个范式框架，其中包括了可扩展的范式取向、对现有 UX 方法的提升和跨学科方法的采用（人因工程、工程心理学等人因科学方法），从而有助于今后进一步充实新方法。

3.2 节将详细阐述“UX 3.0”范式框架中的五大类方法。

## 3.2 “UX 3.0”范式的方法体系分析

### 3.2.1 生态化体验

在移动互联网时代的 UX 第二阶段，体验生态系统开始形成（董建明、傅利民、饶培伦等，2016； Xu，2014，2017），智能时代的 AI、大数据、普适计算、云计算、环境智能、智联网等技术进一步拓展了用户的体验生态系统。目前，用户的体验不再局限于他们在某一时刻对某单一产品中的某个交互接触点上所产生的体验，而是取决于一个动态化、流程式的生态化体验。

研究者已经提出了一些针对生态化体验的初步概念。例如，Klocek（2012）提出了一种多层次结构模型来表征 B2B 软件环境中的 UX 生态，该模型从用户界面设计开始，逐步深入到人机交互模式和业务流程，直到组织文化转变的顶层。Xu（2012，2014）从体验生态角度提出了“无缝衔接 UX”(unified experience)概念和方法，提倡除了提高人机交互可用性，业务流程再造和优化等活动必须是复杂领域 UX 优化的基础。Xu，Furie et al.,（2016, 2019）进一步提出了一个集交互设计、流程优化、系统集成化和系统智能化的系统化人因工程新方法（interaction, process, integration, and intelligence, IPII），该方法通过业务流程再造和优化、人机交互设计、跨系统集成优化以及人机系统智能化的系统化方法，在智能数字化供应链领域提供了一个优化的体验解决方案。辛向阳（2019)认为我们需要从 UX 过渡到体验设计的新范式，设计应该从关注用户生活的手段转而关注用户生命经历的一个过程，将体验经历作为设计对象。而基于跨渠道全域体验和整体体验的全面体验管理策略也体现了生态化体验的思维（黄峰等等，2022）。针对智能时代的人-智能技术-社会子系统之间的交互，许为（2022c）提出了一个智能社会技术系统的概念框架，认为用户在各种智能生态形态（智能交通、智能家居、智能工厂、智能医疗等）中所获得的体验必然受到宏观社会技术环境中各种因素（组织、文化等）的影响。

表 3 系统概括了“UX 3.0”范式中“生态化体验”方法的四种范式取向、生态构成、对“UX 3.0”方法的新要求、部分方法实例。

表 3　生态体验化的范式取向、方法新需求及方法

| 生态化体验的范式取向 | 体验生态构成 | 对“UX 3.0”方法的新要 | “UX 3.0”方法（部分） |
|---|---|---|---|
| **跨产品全生命周期的体验生态链** | UX 不仅仅关注于产品开发阶段针对人机界面的单一用户接触点；用户获取的体验最终来自多触点、多渠道的全流程体验，包括来自跨用户产品需求定义、产品开发、品牌推广、市场投放、用户采购、安装、使用、用户支持、产品更新、退出市场等产品全生命周期各阶段的“体验生态链” | 优化跨产品全生命周期“体验生态链”内各阶段和用户接触点的交互和体验设计，提供端到端的整体 UX | 端到端 UX（需求、设计、测试）方法，用户旅行图，UX 路线图等（Xu，2014） |
| **跨 X 的横向体验生态圈** | UX 不仅仅关注用户对单一产品的体验；基于智联网、大数据、移动技术、云计算、普适计算、智能人机交互等技术，用户获取的体验来自跨平台（如操作系统）、跨设备（如桌面、手机、可戴设备）、跨服务和内容（如购物、旅游、社交媒体、娱乐）等构成的“横向体验生态圈” | 优化“横向体验生态圈”内各部分之间的系统化有效整合设计，提供无缝连接、一致性整体 UX | 端到端 UX（需求、设计、测试）方法，UX 路线图，体验最佳落地区等 |
| **跨系统架构的纵向体验生态层** | UX 不仅仅关注于交互产品的用户界面；用户获取的体验来自前端（如智能人机界面）、中端（如业务逻辑和流程）、后端（如数据库、数据质量和整合、内部部署或云技术或混合云）的跨产品系统架构“纵向体验生态层” | 优化“跨系统架构的纵向体验生态层”内各层面以及跨层面的整合设计 | UX 架构设计，集交互、流程、整合和智能（IPII）的系统化人因工程方法（Xu, Furie et al.，2019） |
| **智能社会技术系统的体验生态环境** | UX 不仅仅关注用户与产品交互中的直接体验；用户获取的体验还来自于各类智能社会形态（如智能城市、智能交通、智能工厂、智能社区）所存在的社会、文化、组织等因素的影响以及这些因素与智能技术之间交互所构成的宏观“智能社会技术系统体验生态环境” | 优化宏观“智能社会技术系统体验生态圈”内各子系统之间的交互设计 | 智能社会技术系统方法（许为，2022c） |

从表 3 可见，智能时代的生态化体验包括：覆盖产品全生命周期的“体验生态链”，跨平台、设备、服务和内容的“横向体验生态圈”，跨系统架构的“纵向体验生态层”，以及宏观智能社会技术系统体验内的“生态环境”，因此，只有系统化的 UX 方法才能为这些体验生态系统内的用户提供无缝衔接、端到端的整体 UX。

### 3.2.2 创新赋能体验

创新设计已经是当前社会经济发展的动力之一。随着产品与服务同质化趋势日益严重，快速发展的技术和不断提高的用户期望，创新设计显得越来越重要。同时，“体验”价值日益凸显，逐渐成为企业或品牌构建持续性、差异化、高价值竞争优势的关键要素。

创新本质上是一种持续地将用户体验（用户需求、使用场景等)与技术的不断调整达到最佳的人机匹配的过程，使技术有用、易学、易用，从而为人创造一种新体验的生活和工作方式（Evans，Buckland & Lefer，2006；许为，2019a）。这种“实用性”创新过程本质上就是基于技术发明的 UX 驱动式创新, 这正是“以用户为中心设计”所倡导的理念。

越来越多的研究强调 UX 对创新的贡献。罗仕鉴（2020）认为进入数据智能时代，群智创新应运而生。以往的创新方式各有利弊，但是他们的一个共同点是没有强调 UX 在创新设计中的作用，人们过分强调技术对创新设计的驱动，忽视了 UX 对创新的作用，导致比较高的失败率（Yang et al., 2018; Debruyne，2014；李四达，2017)。

许为（2019）提出了一个体验驱动式创新的“三因素”概念模型。该模型强调成功的创新设计需要充分考虑用户、技术和环境三因素之间的权衡；创新本质上是一种 UX 驱动的过程，即它是从用户需求出发，通过提炼和洞察用户需求、用户行为、使用场景，权衡技术和环境等因素，发现或预测创新的体验，满足用户需求和体验，从而达到体验驱动创新的目的。表 4 概括了“UX 3.0”范式中“创新赋能体验”方法的五种范式取向、方法特征以及部分方法描述（详见许为，2019a）。

表 4　体验驱动创新的范式取向、特征及方法

| 创新赋能体验的范式取向 | “UX 3.0”方法特征 | “UX 3.0”方法（部分） |
|---|---|---|
| **用户需求** | 基于用户痛点 | 采用传统用户研究方法或 AI、大数据等新技术手段，分析洞察同类产品中的共同用户痛点，找到解决用户痛点的独特体验解决方案 |
| | 基于潜在用户需求和使用场景 | 采用传统用户研究方法或 AI、大数据等新技术手段，挖掘或预测潜在的（尚未发现或实现）、有价值的用户需求和使用场景 |
| | 基于用户差异化体验 | 采用传统用户研究方法或 AI、大数据等新技术手段，挖掘适合有效使用场景的关键体验,即同类产品不具备的、符合用户需求的新型体验 |
| **人机交互技术** | 基于人机交互新技术 | 利用人机交互新技术，开发易学、易用、有价值的新型体验 |
| | 基于现有人机交互技术 | 利用现有人机交互技术，发现符合用户需求的使用场景和最佳落地体验（例如，采用现有单通道人机交互技术，通过多模态人机交互手段来解决用户痛点） |
| **数智技术** | 基于在线用户行为数据 | 利用 AI、大数据等技术，利用实时用户行为等数据，建模和分类用户特征模式，推出符合个性化需求的功能、服务和内容 |
| | 基于上下文场景 | 利用 AI、大数据等技术，通过对实时用户行为、上下文场景数据建模，预测用户需求和使用场景 |
| **人智协同** | 基于动态化人机功能分配 | 根据智能系统学习能力的变化，动态调整人机功能和任务分配，达到最佳人机匹配和人机系统效率 |
| | 基于人机混合增强智能 | 将人类智能引入智能系统中，形成人在回路（或“脑在回路”）、以人为中心、人机智能互补、更强大的人机混合增强智能 |
| | 基于人机组队式合作 | 在感知、认知、执行层面上，实现人智能系统之间信息、目标、任务、执行、决策的双向分享，形成有效的人-智能体合作伙伴，提升人机系统的整体绩效和体验 |
| **端到端的整体体验** | 基于社会技术系统 | 在社会技术系统大环境中，采用“UX 3.0” 范式方法，为用户提供端到端的整体体验解决方案（如创新的服务设计） |

表 4 所列的创新方法都体现了用户与系统、产品或服务在不同范围和层面上的交互接触点，UX 产生于这些交互接触中，因此，这些方法均体现了体验驱动创新的思路。这些创新方法充分利用了新技术（如 AI,大数据，人机交互），通过实时用户分析、动态建模等方法，超越传统 UX 方法。另外，UX 驱动式创新不完全依赖于新技术发明，采用现有技术或者基于以人为中心设计等方法，人们同样可以完成创新，例如，基于用户需求的创新方法，基于现有人机界面技术的创新方法，基于端到端的整体 UX 解决方案。最后，这些创新设计方法具有可操作性。随着体验创新实践的开展，UX 专业人员要与其他学科团队密切合作，进一步细化和完善这些创新方法。

### 3.2.3 AI 赋能体验

AI 技术对 UX 方法带来了革命性的影响，越来越多的 AI 技术被应用在 UX 研究和应用中。AI 赋能体验可以从两个方面来实现。

从 UX 专业人员角度看，AI 技术提升了 UX 专业人员的工作效率和体验设计质量。例如，在用户研究中，采用基于大数据和机器学习方法来构建数字化用户画像，为实时个性化设计和体验提供基础。在人机界面原型化设计中，将 AI 功能（如智能搜索）作为一种“新型设计材料”在用户界面原型化中构建支持 UX 的功能（Dove et al., 2017）；采用生成式 AI 设计工具，支持用户界面原型化。在可用性测试中，采用智能化工具提高数据采集和分析的效率；生成式 AI 工具(如 ChatGPT，Bard) 今后不仅能快速生成用户界面和人机交互代码，也可能会智能化地进行错误检测（如可达性检测）。这些方法有效地提高了 UX 实践的效率，提升了解决方案的 UX。从用户角度看，AI 赋能体验帮助提升用户体验， ChatAPT、Bard 就是很好的实例。

表 5 概括了“UX 3.0” 范式中“AI 赋能体验”方法的范式取向和部分方法实例(详见 Xu, 2023)。

表 5　AI 赋能体验的范式取向和方法

| AI 赋能体验的范式取向 | “UX 3.0”方法（部分） | 方法目的 |
|---|---|---|
| AI 赋能用户研究 | 用户需求采集、洞察和分析，用户访谈语音内容智能化分析，用户数据智能质化分析，数智化用户画像，数智化客户旅程图 | • 提升 UX 方法<br>• 提高 UX 活动效率<br>• 优化体验设计<br>• 增强用户实时体验 |
| AI 赋能用户界面原型化 | AI 生成式 UI 设计（智能化界面图形生成）(Li et al., 2020；ChatGPT、Bard、Adobe Firefly 等 AI 生成式工具)，基于“AI 作为设计新材料”理念的智能化体验设计，人智协同 UI 设计（De Peuter et al., 2021) | |
| AI 赋能体验测试 | 智能化体验测试，用户在线数据分析，多模态（眼动、表情、脑电等）数据收集和分析(Chromik et al., 2020)，视觉搜索绩效建模 (Yuan et al., 2020). | |
| AI 赋能用户实时体验 | 基于实时数据（用户行为、上下文等）建模的个性化需求推理、功能和内容推送，对话式智能助手系统（ChatGPT、Bard 等），智能推荐系统，智能语音助手 | |

从表 5 可知，在 UX 流程的主要阶段（用户研究、界面原型化、体验测试等），AI 技术可以提供更有效的方法或者提升现有 UX 方法，可以为用户提供更丰富的体验。总的来说，这些方法目前还没有完全成熟。例如，针对用户界面原型化设计的 AI 设计辅助工具，它重视外观的新颖性，但缺乏对 UX 设计本身的关注，无法为 UX 设计人员提供最佳的 UX（如构建基于有效业务流程和交互架构的用户界面）。并且，一些基于 AI 技术的原型制作工具的目标用户大多是软件开发人员，导致这些工具不满足 UX 设计专业人员以及 UX 设计流程的特殊需求（Sun et al., 2020）。从长远来看，AI 技术可以作为 UX 设计的辅助工具助手来提升 UX 设计人员的工作效率，但是可能无法完全取代富有创造力的 UX 设计。UX 专业人员需要主动与 AI 专业人员合作，根据 UX 活动的需求来优化这些方法。

### 3.2.4 人智交互体验

传统人机交互（HCI）是 PC 时代形成的跨学科领域,它研究人-非智能计算系统之间的交互。基于智能技术，传统人机交互正在转变到智能人机交互，即人智交互（人与智能系统的交互）。这种变化既带来了更加丰富的体验，同时也对智能时代的 UX 方法提出了新要求(Ozmen Garibay, Winslow,...& Xu, 2023)。

基于 AI 技术的智能系统（体）具有独特的自主化特征，通过开发可以使智能体具备一些类似人类的认知特征（如感知、自我学习、自主执行）（许为，2020）。一方面，这些新特征对开发自然有效的人机交互和体验提供了强大的技术支持，例如，语音，脑机界面，虚拟现实/虚实混合，情感交互等；另一方面，这些新特征也带来了独特的体验问题，以往基于非智能计算系统的 UX 方法无法有效地解决这些新问题。表 6 概括了智能技术带来的人智交互新特点、人智交互对 UX 方法的新要求以及 UX 方法描述。

表 6 “UX 3.0” 范式的人智交互体验方法

| 人智交互体验的范式取向 | 现有人机交互特点 | 人智交互新特点 | 人智交互对 UX 方法的新要求 | “UX 3.0”方法（部分） |
|---|---|---|---|---|
| **智能系统演化行为优化** | • 可预期的机器行为<br>• 设计依赖于预期的机器行为，但是无法处理异常场景 | • 智能系统可展示不确定机器行为，其行为可随算法学习而演化(Rahwan et al., 2019)<br>• 智能机器行为可导致潜在有偏见的输出，影响 UX (Abbas et al., 2022)<br>• 智能自主化能力（如自学习、自执行）可处理一些设计无法预料的场景 (Xu, 2019) | • 减少算法偏见，优化机器行为演化<br>• 支持人类可控 AI 的体验<br>• 防范用户对 AI 过度信任的体验<br>• 提升用户监控、预测应急状态以及处理异常场景的能力 | • 采用用户参与、迭代式原型设计和测试，优化算法数据收集、训练和测试<br>• 收集用户反馈，支持机器行为演化的持续优化<br>• 有效的人机交互和界面设计 |
| **人机混合智能** | • 单一的人类智能<br>• 非智能系统不具备机器智能<br>• 体验设计无法利用机器智能 | • 利用人机混合智能（人类智能和机器智能），可提升系统整体绩效和体验 | • 借助机器的类人智能，支持人类体验<br>• 通过“人在回路”的人机混合智能，提升人智交互中的体验 | • 通过人类用户与智能机器的功能分配，优化人机关系和体验<br>• 通过人类智能增强方法，提升体验（Raisamo et al., 2019） |
| **智能系统输出可解释性** | • 系统输出用户界面的可用性设计 | • AI “黑箱效应”可导致智能系统输出对用户晦涩难懂（AI 可解释性问题）（杨强等， 2022）<br>• 影响用户信任度和体验 | • 以用户为中心的可解释 AI<br>• 验证可解释 AI 设计的有效性 | • 可视化人机界面设计<br>• 用户参与的设计和 UX 测试<br>• 基于可解释性设计的可理解性（被用户） |
| **自然式智能人机交互** | • 图形用户界面可用性设计<br>• 基于“简单属性”、“精准输入”的人机交互（易鑫，喻纯，史元春，2018） | • 基于智能技术的自然人机交互（如语音输入）<br>• 情境化、模糊推理的人机交互<br>• 智能体主动式交互（如基于用户意图、情感识别）的智能识别和推演<br>• 社会和情感交互 | • 更加自然和可用的人机交互<br>• 智能产品的智能化（自主化）水平差异对体验设计的新要求<br>• 智能人机交互设计标准化<br>• 支持智能化人机界面原型化设计<br>• 支持社会、情感交互 | • 用户状态和意图识别建模方法<br>• 基于智能化水平的差异性体验设计方法<br>• 智能人机交互设计新范式<br>• 智能人机交互设计标准（Amershi et al., 2019）<br>• 智能交互人机界面原型化方法<br>• 社会、情感交互体验设计方法 |
| **伦理化 AI 需求** | • 基于对产品可用性、功能、安全性等一般用户需求的体验 | • 基于用户新需求（如隐私、公平、道德、技能增长、决策权）的体验 | • 智能系统应该反映用户新需求和相应的体验 | • 将用户新需求整合到数据收集、算法训练、测试和优化方法中（如 UX 迭代式设计和测试）<br>• 基于行为科学等跨学科方法的伦理化 AI 设计 |

显而易见，UX 专业人员需要清楚地认识到智能技术对人机交互设计和体验带来的革命性变化。如表 6 所示，来自跨学科专业的 UX 人员具备行为科学、人因科学等领域方法的专业知识，这些领域知识和领域方法的特征为提供优化的人智交互体验方案提供了帮助，同时可以形成与 AI 专业人员的学科互补性，有助于促进协同合作和提升 UX 方法。

例如，针对 AI 可解释性问题，UX 专业人员可以利用工程心理学和人因工程等人因科学方法，研究探索式、自然式、交互式解释来设计可解释人机界面(Abdul et al., 2018)，开发“以人为中心的可解释 AI”解决方案。AI 专业人员通常没有受过行为科学训练，针对智能机器行为的非确定性问题，UX 人员可以利用早期界面原型化和 UX 测评的迭代式方法，从数据、算法学习、测试方面来减小或避免算法偏差(Buolamwini & Gebru, 2018)，即“以人为中心” 机器学习、交互式机器学习方法(Kaluarachchi et al., 2021)。另外，智能系统丰富的应用场景和用户需求需要有效的人机界面设

计新范式（范俊君, 田丰, 杜一等, 2018），而现有人机交互方式局限于感知通道有限、交互宽带不足、交互方式不自然等问题, 因此 UX 领域可以在开拓界面设计新范式等方面作出贡献。

### 3.2.5 人智协同合作体验

智能技术带来了智能时代的新型人机关系（许为，葛列众，2020）。智能技术的自主化特征赋予人机系统中机器新的角色。在非智能时代，人类操作基于计算技术，机器充当辅助工具角色。智能时代的人与智能系统的交互本质上是人与自主智能体之间的交互。随着智能技术的提升,自主智能体有可能从一种支持人类操作的辅助工具角色发展成为与人类操作员共同合作的团队队友，扮演“辅助工具 + 人机合作队友”的双重新角色（许为，2020；Xu，2020）。由此，智能时代的人机关系正在演变成为团队队友关系，形成一种“人机组队”（human-machine teaming）式合作关系，在一定程度上可以实现类似于人-人团队队友之间的协同合作（Brill, Cummings et al., 2018）。人类操作员与自主智能体两者之间的这种协同合作具有双向主动的、分享的、互补的、可预测等特征（许为，葛列众，2020）。随着 AI 技术的发展，未来智能系统将更加具备这些协同合作的特征。

在智能时代，基于人智协同合作式的新型人机关系进一步对人机界面和体验设计提出了新要求。传统的非智能人机交互是基于“刺激-反应”理念的“指令顺序”式单向式交互（Farooq & Grudin, 2016),UX 专业人员需要改变原有的思维方式。根据我们提出的人智组队（human-AI teaming）人因工程概念模型，在智能人机交互中，智能体也是一个认知体，人智交互系统可以视为一个协同认知系统，从而智能体与人类用户认知体可以开展类似人-人团队式的协同合作（Hollnagel & Woods, 2005；许为，2022a）。因此，这种人智协同合作之间的体验设计需要有效的人智合作式界面，从而支持这种人智双向式的情景意识和心理模型分享、人机互信、人机控制分享、人智情感交互、用户意图识别等人智协同合作活动。

智能时代人机关系本质上的新变化为提升人智交互的体验提供了更多的可能性，也为人智协同合作语境下的体验设计提出了新要求。人智交互与人智协同合作有着密切的联系，人智协同合作基于人智交互，又超越人智交互。目前，ChatGPT-4、Bard 等 AI 语言大模型系统的应用已经开始显现出人智协同合作的思路强调。从“以人为中心设计”的理念看，AI 只是一个副机长（copilot），协助用户的更高效产出 （如 Microsoft Office 365 Copilot 系统），但是，目前这些系统远没有达到这里所定义的“人机组队”式合作关系。UX 专业人员需要充分理解这种新兴的人机关系，探索如何利用这种人智协同合作关系来进一步提升用户使用智能系统的体验。

人智协同研究已经在工程心理学、人因工程、HCI、AI 等领域展开（NASEM，2021）。表 7 概括了人智协同的新特点、对“UX 3.0” 范式中“人智协同体验”方法的新要求以及部分 UX 方法。

表 7　人智协同合作体验的范式取向、新特点及对 UX 方法的新要求（部分）

| 人智协同合作体验的范式取向 | 人智协同合作的新特点 | 人智协同合作对 UX 方法的新要求 | “UX 3.0”方法（部分） |
|---|---|---|---|
| **用户、智能体双向式感知和控制** | • 人类用户和智能体两个认知体会产生双向的“交互”(许为，2022a)<br>• 人智之间拥有双向的信任、情景意识以及决策分享（人应该拥有最终控制权）<br>• 人智双方均可主动地启动任务和行动<br>• 人智双方均可预测对方的行为状态 | • 人智之间的人机共信<br>• 人智之间的分享式心理模型、情景意识、决策和控制 | • 感知、理解、预测人智协同状态的情境意识建模方法<br>• 人智协作的体验需求定义、模型和方法 |
| **人智协同式动态化功能分配和优化** | • 智能系统具备学习能力，人智系统优化需要基于人智之间的“动态化”任务分析和功能分配<br>• 人智团队可以设置任务目标，长期合作<br>• 基于人智功能分配的权限转换和控制共享 | • 人智协同语境下的任务分析、人机功能分配（不同于传统方法）<br>• 人智协同“动态化”语境下，利用人机智能互补性，达到最佳人机混合智能的体验 | • 人智协同合作语境下的任务分析、人机功能分配方法 |
| **合作式认知人机界面** | • 作为团队队员，人类用户和智能体开展协同合作 | • 支持人智协同合作的人机界面<br>• 支持人智协同合作的优化信息集成、可解释性和可理解性<br>• 应急状态下，支持人智之间控制权快速有效切换的人机界面 | • 合作式认知界面设计新范式和新隐喻<br>• 合作式认知界面设计原型化<br>• 支持人智之间控制权快速有效切换的合作式人机界面 |
| **人智协同团队绩效和体验测评** | • 人类用户和智能体可以共同学习和成长,影响体验<br>• 智能系统输出的不确定性和不可预测性会影响体验 | • 人智协同语境下的的体验以及体验演化<br>• “动态化”复杂场景中的人智团队绩效和体验测评<br>• 人-智能体共同学习和成长的体验测评 | • 人智协同合作的团队绩效和用户体验测评方法系统 |

“UX 3.0”范式框架中将“人智交互体验”与“人智协同合作体验”两者区分是基于以下考虑：首先，如表 6 和表 7 所概括，人智交互与人智协同合作具有各自的新特点、对体验方法的新要求以及相应的“UX 3.0”方法，而且人智协同合作体验对智能体、人智能系统人机界面等方面的设计提出了更高的要求；其次，目前的研究和应用主要集中在人智交互方面，人智协同合作方面尚不成熟。如果将两者置于一个“人与智能系统（智能体）整合空间”中，随着研究和应用的推进，我们今后会对该“整合空间体”中人智交互与人智协同合作之间的本质关系有一个更加全面的认识；最后，在目前阶段，区分这两种体验有助于促进人们对人智协同合作以及相应体验中一些特殊人因科学问题的足够关注，从而推动针对性的研究和应用。随着研究和应用的深入开展，我们今后可以对这两种体验做出全面的科学评价。

针对表 7 中初步提出的一些方法和思路，UX 专业人员要主动参与与其他学科的合作，在相关学科（AI、工程心理学、HCI、人因工程等）的指导下，根据 UX 领域特点，探索人智协同合作语境下的

体验设计，例如，探索人机（智）情景意识共享、人机互信、人机心理模型共享、人机决策共享、合作式人机界面、人智协同合作体验评估等新问题，这些问题也是今后智能时代 UX 领域的重要研究课题。

### 3.3 “UX 3.0”范式新特征及意义

基于以上对“UX 3.0” 范式和方法体系的阐述和分析，在表 1 基础上，图 2 和表 8 从 UX 范式的角度进一步比较了 UX 三个发展阶段之间的跨时代特征。

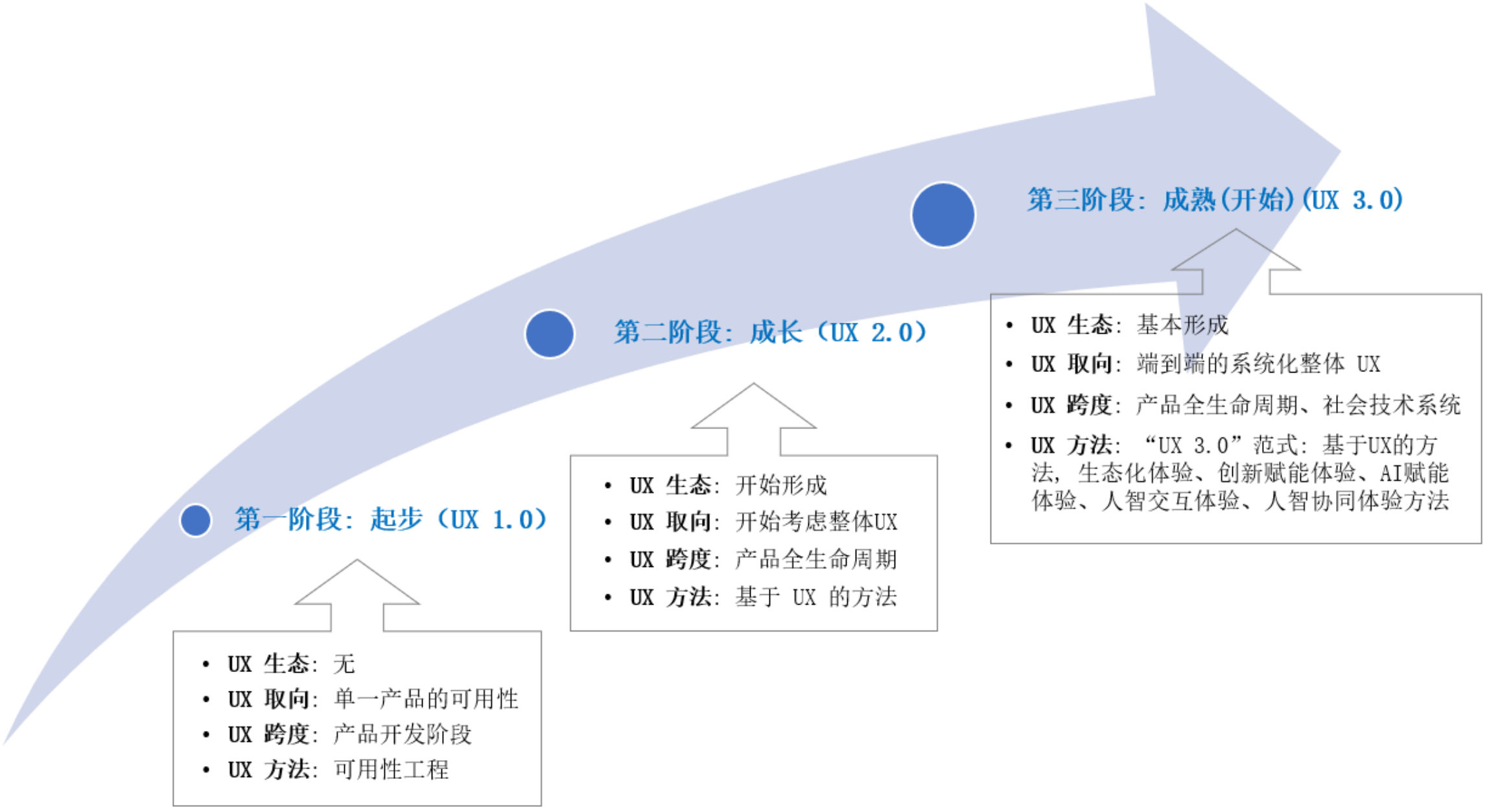

图 2 UX 范式演进的三阶段

表 8　UX 范式跨时代特征比较

| 范式特征 | 第一阶段：UX 1.0<br>（1980 年代后 - ～2007 年） | 第二阶段：UX 2.0<br>（～2007 年 - ～2015 年） | 第三阶段：UX 3.0<br>（～2015 年 - ） |
|---|---|---|---|
| **技术时代** | 个人电脑/互联网时代 | 移动互联网时代 | 智能时代 |
| **UX 范式成熟度** | 起步 | 成长 | 成熟（开始） |
| **UX 生态** | 无 | 开始形成 | 基本形成 |
| **UX 取向** | 单一交互式产品的可用性 | 开始考虑整体 UX（包括可用性），开始考虑基于端到端 UX 的整体体验 | 端到端的系统化整体 UX：<br>+ 生态化体验 + 创新赋能体验 + AI 赋能体验 + 人智交互体验 + 人智协同合作体验 |
| **UX 跨度** | 产品开发阶段 | 产品全生命周期（开发前、开发、开发后） | 产品全生命周期 + 宏观智能社会技术系统 |
| **UX 方法** | "UX 1.0"范式：可用性工程（Nielson, 1993）：<br>基于可用性的方法。包括传统用户调查问卷、访谈，界面原型化，可用性测试等 | "UX 2.0"范式：基于 UX 的方法（例如，许为，2017）：<br>包括可用性方法，端到端的整体体验设计，技术赋能方法（如用户大数据分析）等 | "UX 3.0"范式：<br>基于 UX 的方法 + 生态化体验方法 + 创新赋能体验方法 + AI 赋能体验方法 + 人智交互体验方法 + 人智协同合作体验方法等 |
| **团队合作** | 可用性专业人员，UI 开发人员 | 精细化分工和扩展的 UX 团队<br>• 用户研究<br>• 业务流程再造<br>• UI、视觉设计，内容设计<br>• 前端开发<br>• 可用性测试<br>• 用户支持等 | 强化的跨学科合作团队<br>• 精细化分工和扩展的 UX 团队<br>• AI、数据工程师，系统架构师等<br>• 人因科学（人因工程、工程心理学、人机交互等）团队<br>• 所有项目团队成员以及相关利益者均是"UX 设计者" |

由图 2 和表 8 可见，根据评价 UX 范式成熟度的各个维度（如 UX 生态、UX 取向、UX 跨度、UX 方法），40 多年来，UX 范式的演进也清楚地体现出如表 1 所定义的三个阶段，UX 范式的成熟度从第一阶段的"起步"到第二阶段的"成长"，再到目前第三阶段的"成熟"（开始）。 而且，UX 范式演进的三阶段之间存在本质上的差别，这种"UX 范式三阶段"的划分也符合 UX 实践成熟度的发展规律（葛列众，许为，2020：见第 11 章；陶嵘, 黄峰，2012）。

"UX 范式三阶段"概念和"UX 3.0"范式的提出对 UX 研究和应用具有重要意义。首先，有助于 UX 专业人员清醒地认识到，进入智能时代，UX 行业进入了实践深水区，智能时代的 UX 研究和应用对 UX 方法论提出了新要求，现有 UX 方法不能完全满足智能时代的 UX 实践，我们需要提升现有 UX 方法。其次，有助于促进 UX 专业人员开展必要的研究和应用来进一步提升现有 UX 方法。进入 UX 第三阶段，采用和完善"UX 3.0"范式和方法体系也是今后 UX 研究和应用的重要课题之一。最后，有助于 UX 专业人员主动强化与其他相关学科的协同合作（见表 8），取长补短，进一步提升现有 UX 方

法。UX 本身是一个通过跨学科协同合作所产生的新兴领域，40 多年前 UX 领域的产生和方法也是来自于其他学科的推动，智能时代的 UX 研究和应用更加需要其他学科的支持，这对于一门年轻的、入门门槛相对较低的领域来说尤其重要。

## 4 “UX 3.0”范式研究和应用展望

本文提出的“UX 3.0”范式是基于以往的 UX 研究和应用，一些方法已经具备可操作性，但是许多方法还处于不成熟阶段，需要进一步完善，同时还需要充实新的方法。针对下一步“UX 3.0” 范式的研究和应用，我们需要考虑一下几方面的工作。

首先，建立设计新思维。UX 专业人员需要将智能时代和智能技术应用视为提升 UX 方法论的重要机遇，智能技术不仅仅是工具，UX 专业人员需要转变设计思维，这样才能提供符合智能时代需求的有效体验解决方案。例如，“创新赋能体验”方法中的体验驱动创新思维、“人智协同合作体验”方法中的新型人机关系、“智能社会技术系统”范式取向的宏观视野、“跨产品全生命周期体验生态”的设计跨度、系统化的整体 UX。

其次，开展对“UX 3.0”范式和方法的研究以及应用。针对“UX 3.0” 范式和方法的研究，不仅涉及到与其他相关学科的协同合作，而且也需要进一步对这些方法的具体操作、量化测评和指标体系、流程标准化、辅助工具等方面开展工作，同时通过 UX 实践进一步优化这些方法。

第三，提高 UX 专业人员素质。在第一、二阶段，UX 领域基本是一个入门门槛低、各学科和行业融合的实践浅水区，进入智能时代的 UX 实践深水区（第三阶段），UX 行业人才需求呈现快速增长态势，同时 “UX 3.0”方法的实践也需要更加专业化的知识和技能。在职 UX 专业人员需要提升智能技术、系统化方法、人因科学（如工程心理学、人因工程）等方面的知识；针对 UX 专业在校学生或对 UX 领域感兴趣的在校学生，高校应该开设“人智交互”、“以人为中心 AI”、“UX + AI”等课程；另外，通过科研和研究生培养开展针对“UX 3.0” 范式和方法的研究，培养满足智能时代需求的高层次 UX 专业人才。

第四，强化跨学科协同合作。完善“UX 3.0”范式和方法的工作已经超出了现有 UX 领域的知识体系范围，需要跨学科领域的通力合作，这些学科包括工程心理学、人因工程、HCI、AI、计算机等。同时“UX 3.0” 范式和方法的实践也需要这些跨学科团队的支持。

最后，优化智能产品开发流程和组织环境。“UX 3.0” 范式的有效应用也取决于这些 UX 方法是否能够有效地被整合到产品开发流程、方法以及环境中。在智能产品开发项目层面上，最大限度地实现跨学科协作，通过建立多学科项目团队和采用跨学科方法，基于“以用户为中心”理念来优化现有的智能产品开发流程。在智能产品开发企业和组织层面上，培育基于“以用户为中心”理念的组织文化，制定基于人智交互的开发标准指南，优化开发人力资源，建立高效的跨学科 UX 合作团队，在实践中不断优化“UX 3.0” 范式和方法。

## 5 小结

纵观 40 多年的发展历史，UX 呈现出明显的阶段性发展特征。新兴技术和 UX 实践推动了 UX 范式的发展，这种发展提升了 UX 研究和应用，也推动了 UX 领域的成长。目前，UX 领域正在迈向智能时代,智能时代对 UX 范式提出了一系列新要求。

为了满足智能时代对 UX 范式的新要求，本文提出了一个智能时代“UX 3.0”范式的框架，该框架包括生态化体验、创新赋能体验、AI 赋能体验、人智交互体验、人智协同合作体验等五大类 UX 方法体系，该框架也定义了各类方法中所包括的各种范式取向以及具体的 UX 方法。

“UX 3.0”范式有助于 UX 专业人员清醒地认识到，智能时代对 UX 方法论提出了新要求，现有 UX 方法不能完全满足智能时代 UX 研究和应用的需要，我们需要提升现有的 UX 方法，这也是今后 UX 研究和应用的重要课题之一。UX 领域需要在研究与应用中通过跨学科合作等方法进一步提升和完善智能时代的 UX 范式。

## 参考文献

葛列众，许为. (2020). *用户体验：理论与实践*. 北京,中国人民大学出版社.

葛列众, 许为，宋晓蕾. (2022). *工程心理学(第二版).* 北京, 中国人民大学出版社.

吕超，朱郑州. (2018). 一个基于用户画像的商品推荐算法的设计与应用. *中国科技论文在线*, *11*(4), 339-347.

李四达.(2017).*交互与服务设计*. 北京: 清华大学出版社.

易鑫, 喻纯, 史元春. (2018). 普适计算环境中用户意图推理的 Bayes 方法.*中国科学：信息科学*.

董建明，傅利民，饶培伦，Stephanidis，C.& Salventy, G.（2016）. *人机交互：以用户为中心的设计和评估（第 5 版）*. 北京：清华大学出版社.

董建明，傅利民，饶培伦，Stephanidis，C.& Salventy, G.（2021）. 人机交互：以用户为中心的设计和评估（第 6 版）北京：清华大学出版社.

谭浩, 尤作, 彭盛兰. (2020). 大数据驱动的用户体验设计综述. *包装工程*, *41*(2), 7-12.

陶嵘, 黄峰. (2012). 用户体验成熟度研究. UXPA 中国研究报告.

许为. (2003). 以用户为中心设计：人机工效学的机遇和挑战. *人类工效学*, *9*(4), 8-11.

许为. (2005). 人-计算机交互作用研究和应用新思路的探讨. *人类工效学*, 11(4), 37-40.

许为. (2017). 再论以用户为中心的设计: 新挑战和新机遇. *人类工效学*, *23*(1), 82-86.

许为. (2019a). 三论以用户为中心的设计：智能时代的用户体验和创新设计.*应用心理学*, *25*(1), 3-17.

许为. (2019b). 四论以用户为中心的设计: 以人为中心的人工智能. *应用心理学*, *25*(4),291-305.

许为. (2020). 五论以用户为中心的设计: 从自动化到智能时代的自主化以及自动驾驶车. *应用心理*

*学*, 26(2), 108-
128.
许为. (2022a). 六论以用户为中心的设计: 智能人机交互的人因工程途径. *应用心理学. 28(3), 191-209*.
许为. (2022b). 七论以用户为中心的设计: 从自动化到智能化飞机驾驶舱. *应用心理学. 28(4), 291-313*.
许为. (2022c). 八论以用户为中心的设计: 智能社会技术系统，*应用心理学*. 28(5), 387-401
许为, 葛列众.(2018). 人因学的新取向（主编特邀）.*心理科学进展*, *26*(9),1521-1534.
许为, 葛列众. (2020). 智能时代的工程心理学（主编特邀）.*心理科学进展*, 28(9), 1409-1425.
许为，葛列众，高在峰.(2021). 人-AI 交互: 实现"以人为中心 AI"理念的跨学科新领域. *智能系统学报*, 16(4), 604-621.
许为，高在峰, 葛列众.(2022). 智能时代的人因科学：研究范式和重点.
https://arxiv.org/abs/2208.12396
辛向阳. (2019). 从用户体验到体验设计. *包装工程艺术版*, *40*(8), 60-67.
罗仕鉴. (2020). 群智创新: 人工智能 2.0 时代的新兴创新范式. *包装工程*, *41*(6), 50-56.
黄峰,赖祖杰. (2020). 体验思维, 天津, 科学技术出版社
黄峰,黄胜山,苏志国.(2022).全面体验管理 TXM. 北京,中国财政经济出版社
杨强、范力欣、朱军等.（2022）. 可解释人工智能导论，北京,电子工业出版社
范俊君, 田丰, 杜一, 刘正捷, 戴国忠.(2018). 智能时代人机交互的一些思考. 中国科学: 信息科学, 2018, 48(4): 361 - 375.
Abdul, A., Vermeulen, J., Wang, D., Lim, B. Y., & Kankanhalli, M. (2018, April). Trends and trajectories for explainable, accountable and intelligible systems: An hci research agenda. In *Proceedings of the 2018 CHI conference on human factors in computing systems* (pp. 1-18).
Abbas, A. M., Ghauth, K. I., & Ting, C. Y. (2022). User Experience Design Using Machine Learning: A Systematic Review. *IEEE Access*.
Amershi, S., Weld, D., Vorvoreanu, M., Fourney, A., Nushi, B., Collisson, P., ... & Horvitz, E. (2019, May). Guidelines for human-AI interaction. In *Proceedings of the 2019 chi conference on human factors in computing systems* (pp. 1-13).
Budiu，R. & Laubheimer, P. (2018). Intelligent Assistants Have Poor Usability: A User Study of Alexa, Google Assistant, and Siri. Retrieved November 16, 2018, from https://www.nngroup.com/articles/intelligent-assistant-usability/
Buolamwini, J., & Gebru, T. (2018, January). Gender shades: Intersectional accuracy disparities in commercial gender classification. In *Conference on fairness, accountability and* *transparency* (pp. 77-91). PMLR.

Brill, J., Cummings, M. L., Evans III, A. W., Hancock, P. A., Lyons, J. B., & Oden, K. (2018, September). Navigating the advent of human-machine teaming. In *Proceedings of the human actors and ergonomics society annual meeting* (Vol. 62, No. 1, pp. 455-459). Sage CA: Los Angeles, CA: SAGE Publications.

Chromik, M., Lachner, F., & Butz, A. (2020, October). ML for UX?-An Inventory and Predictions on the Use of Machine Learning Techniques for UX Research. In *Proceedings of the 11th Nordic Conference on Human-Computer Interaction: Shaping Experiences, Shaping Society* (pp. 1-11).

Debruyne，M.（2014）. Customer Innovation: Customer-centric Strategy for Enduring Growth. Kogan Page.

De Peuter, S., Oulasvirta, A., & Kaski, S. (2021). Toward AI Assistants That Let Designers Design. *arXiv preprint arXiv:2107.13074*.

Dove, G., Halskov, K., Forlizzi, J., & Zimmerman, J. (2017, May). UX design innovation: Challenges for working with machine learning as a design material. In *Proceedings of the 2017 chi conference on human factors in computing systems* (pp. 278-288).

Evans, H., Buckland, G. and Lefer, D. (2006) *They Made America: From the Steam Engine to the Search Engine: Two Centuries of Innovators*. Back Bay Books

Farooq, U., & Grudin, J. (2016). Human computer integration. *Interactions, 23,* 27−32.

Finstad, K., Xu, W., Kapoor, S., Canakapalli, S., & Gladding, J. (2009). Bridging the gaps between enterprise software and end users. *Interactions*, *16*(2), 10-14.

Ozmen Garibay, O., Winslow, B., Andolina, S., Antona, M., Bodenschatz, A., Coursaris, C., ... & Xu, W. (2023). Six Human-Centered Artificial Intelligence Grand Challenges. *International Journal of Human–Computer Interaction*, 1-47.

Herath, D., & Jayarathne, L. (2018). Intelligent Recommendations for e-Learning Personalization Based on Learner's Learning Activities and Performances. *International Journal of Computer Science and Software Engineering*, *7*(6), 130-137.

Hollnagel, E., & Woods, D. D. (2005). *Joint cognitive systems: Foundations of cognitive systems engineering.* London: CRC Press.
systems. *IEEE Transactions on Human-Machine Systems*, *43*(6), 608-619.

International Standards Organisation （ISO）. (2019). *Ergonomics of Human–System Interaction – Part 220:* Ergonomics of human-system interaction — Part 220: Processes for enabling, executing and assessing human-centred design within organizations. https://www.iso.org/standard/63462.html

Kaluarachchi, T., Reis, A., & Nanayakkara, S. (2021). A review of recent deep learning approaches in human-centered machine learning. *Sensors*, *21*(7), 2514.

Klocek, S. 2012. "Fixing a Broken Experience". *Smashing Magazine.* 2012, Sept., https://www.smashingmagazine.com/2012/09/fixing-broken-user-experience/

Li, F.F. (2018). How to make A.I. that's good for people. *The New York Times*. https://www.nytimes.com/2018/03/07/opinion/artificial-intelligence-human.html.

Li, Y., Kumar, R., Lasecki, W. S., & Hilliges, O. (2020, April). Artificial intelligence for HCI: a modern approach. In *Extended Abstracts of the 2020 CHI conference on human factors in computing systems* (pp. 1-8).

Mcgregor, S. (2023).AI Incidents Database https://incidentdatabase.ai/.

National Academies of Sciences, Engineering, and Medicine (NASEM). (2021). Human-AI Teaming: State-of-the-Art and Research Needs. https://nap.nationalacademies.org/catalog/26355/human-ai-teaming-state-of-the-art-and-research-needs

Nielsen J．(1993). Usability Engineering．Academic Press，

Norman D.A., Draper S.W．(1986). User centered system design ： New perspective on human-computer interaction．New Jersey： Hillsdale．

Preece, J., Rogers, Y., & Sharp, H.（2019). Interaction Design (5th edition), U.K.: John Wiley & Sons Ltd.

Raisamo, R., Rakkolainen, I., Majaranta, P., Salminen, K., Rantala, J., & Farooq, A. (2019). Human augmentation: Past, present and future. *International Journal of Human-Computer Studies*, *131*, 131-143.

Rahwan I, Cebrian M, Obradivuch N. et al. (2019). Machine behavior. Nature, 2019, 568(7753): 477-486.

Quadrana, M., Karatzoglou, A., Hidasi, B., & Cremonesi, P. (2017, August). Personalizing session-based recommendations with hierarchical recurrent neural networks. In *Proceedings of the Eleventh ACM Conference on Recommender Systems* (pp. 130-137).

Salminen, J., Guan, K., Jung, S. G., & Jansen, B. J. (2021). A survey of 15 years of data-driven persona development. *International Journal ofHuman–Computer Interaction*, *37*(18), 1685-1708.

Shneiderman, B., Plaisant, C., Cohen, M., Jacobs, S., & Elmqvist, N. (2016). Design the User Interface (6th edition). England: Pearson.

Shneiderman, B. (2020) Human-centered artificial intelligence: Reliable, safe & trustworthy, *International Journal of Human-Computer Interaction, 36*, 495-504, DOI: 10.1080/10447318.2020.1741118

Stephanidis, C., Salvendy, G., Antona, M., Chen, J. Y., Dong, J., Duffy, V. G., ... & Zhou, J. (2019). Seven HCI grand challenges. *International Journal of Human-Computer Interaction, 35*(14), 1229-1269.

Sun, L. Y., Zhang, Y. Y., Zhou, Z. B., & Zhou, Z. H. (2020). Current situation and development trend of intelligent product design under the background of human-centered AI. *Packaging Engineering*, *41*(02), 1-6.

Wang, L., Gao, R., Váncza, J., Krüger, J., Wang, X. V., Makris, S., & Chryssolouris, G. (2019). Symbiotic human-robot collaborative assembly. *CIRP annals*, *68*(2), 701-726.

Xu, W. & Furie, D. (2016). Designing for unified experience: a new perspective and a case study. Book Chapter (Chapter 5) in *Ergonomics: Challenges, Applications and New perspectives,* edited by Angelina Sparks. Nova Science, ISBN: 978-1-53610-248-2, 137-165.

Xu, W., Furie, D., Mahabhaleshwar, M., Suresh, B., & Chouhan, H. (2019). Applications of an interaction, process, integration and intelligence (IPII) design approach for ergonomics solutions. *Ergonomics*, 62(7), 954-980.

Xu, W.（2012）. User experience design: Beyond user interface design and usability. In Isabel L. Nunes, Ergonomics-A Systems Approach, IntechOpen.

Xu, W. (2014). Enhanced ergonomics approaches for product design: a user experience ecosystem perspective and case studies. *Ergonomics, 57*(1), 34-51.

Xu, W. (2019). Toward human-centered AI: A perspective from human-computer interaction. *Interactions, 26*(4), 42-46.

Xu, W. (2020). From automation to autonomy and autonomous vehicles: Challenges and opportunities for human-computer interaction. *Interactions*, *28*(1), 48-53.

Xu, W. (2021). Maximizing the value of enterprise human-computer interaction standards: strategies and applications. In Waldemar Karwowski & Anna Szopa: *Handbook of Standards and Guidelines in Ergonomics and Human Factors (2nd edition)*, CRC Press, Taylor & Francis.

Xu, W. (2023). AI in HCI Design and User Experience. Constantine Stephanidis & Gavriel Salvendy. BOOK #5 – *Human-Computer Interaction: Interacting in Intelligent Environments.* CRC Press

Xu, W., Dainoff, M.J., Ge, L., Gao,Z.(2021). From human-computer interaction to human-AI Interaction: new challenges and opportunities for enabling human-centered AI. arXiv preprint arXiv:2105.05424 5

Xu, W. & Dainoff, M. (2023). Enabling human-centered AI: A new junction and shared journey between AI and HCI communities. *Interactions, 30*(1), 42-47.

Yang, Q., Banovic, N., & Zimmerman, J. (2018, April). Mapping machine learning advances from hci research to reveal starting places for design innovation. In *Proceedings of the 2018 CHI conference on human factors in computing systems* (pp. 1-11).

Yang, Q., Steinfeld, A., Rosé, C., & Zimmerman, J. (2020, April). Re-examining whether, why, and how human-AI interaction is uniquely difficult to design. In *Proceedings of the 2020 chi conference on human factors in computing systems* (pp. 1-13).

Yuan, A., & Li, Y. (2020, April). Modeling human visual search performance on realistic webpages using analytical and deep learning methods. In *Proceedings of the 2020 CHI conference on human factors in computing systems* (pp. 1-12).